\documentclass[journal]{IEEEtran}

\usepackage{cite}
\usepackage{url}
\usepackage{amsmath,amssymb,amsfonts}
\usepackage{graphicx}
\usepackage{algorithm}
\usepackage{algorithmic}
\usepackage{booktabs}
\usepackage{tikz}
\usepackage{xcolor}
\usetikzlibrary{arrows.meta}
\usepackage{placeins}

\definecolor{leftblue}{RGB}{231,241,250}
\definecolor{rightorange}{RGB}{252,240,226}
\definecolor{neutralgray}{RGB}{242,242,242}
\definecolor{mergegreen}{RGB}{235,245,236}

\begin{document}

\title{A Defense-Aware Explainable AI Framework for Feature-Level Vulnerability Analysis in Optimal Power Flow}

\author{S. M. Monzurul Hoque Chowdhury,
        Qinghua Li,
        and Jianhui Wang%
\thanks{S. M. Monzurul Hoque Chowdhury and Q. Li are with the Department of Electrical Engineering and Computer Science, University of Arkansas, Fayetteville, AR, USA. Their contact emails are sc327@uark.edu and qinghual@uark.edu, respectively.}
\thanks{J. Wang is with the Department of Electrical and Computer Engineering, Southern Methodist University, Dallas, TX, USA. The contact email is jianhui@smu.edu.}
}

\maketitle

\begin{abstract}
Optimal power flow (OPF) is a foundational analysis in power system operation, and inaccuracies in OPF decision-making can result in significant operational and security consequences. With the increasing adoption of machine-learning-based OPF surrogate models for real-time and large-scale applications, power system operation has become more susceptible to cyber threats, including data manipulation and false data injection attacks. Explainable artificial intelligence (XAI), particularly Shapley value–based approaches, has been employed to quantify the influence of input features on OPF outputs. Features exhibiting high influence are commonly interpreted as vulnerable. However, this interpretation neglects the role of attack detection mechanisms in the system, which typically monitor input-to-input relationships to detect anomalous behavior. Consequently, highly influential features may be readily detectable and may not represent the most vulnerable attack surfaces. To address this limitation, we propose a defense-aware explainable AI framework that explicitly incorporates attack detectability alongside operational impact. The framework integrates an OPF surrogate model representing input-to-output dependencies with a vanilla autoencoder representing input-to-input dependencies. Shapley-based explainability is applied to both models to construct a feature-level vulnerability metric. Empirical evaluation on IEEE test systems demonstrates that the proposed framework more accurately identifies security-relevant vulnerabilities than conventional OPF-surrogate-model-only explainability approaches.
\end{abstract}


\section{Introduction}

\subsection{Background and Motivation}

Optimal power flow (OPF) analysis constitutes one of the core components of power system operation that enables system operators to determine economically optimal and technically feasible operating points under network, generation, and operational constraints \cite{wood2012power}. OPF solutions directly affect generator dispatch, voltage regulation, and overall system reliability, making them critical to both planning and real-time operation. As OPF lies at the core of operational decision-making, errors or manipulation in this analysis can significantly disrupt power system stability and lead to severe operational consequences.

With the increasing scale and complexity of modern power systems, machine-learning-based OPF surrogate models have been increasingly explored as computationally efficient alternatives to conventional nonlinear solvers, particularly for real-time analysis and large-scale scenario evaluation \cite{donnot2019deep}. In this work, an OPF surrogate model refers to a machine learning model trained to approximate the mapping between system input measurements (e.g., load demands) and OPF outputs (e.g., generator set-points), without explicitly solving the underlying nonlinear optimization problem. Recent studies on stochastic optimal power flow (SOPF) have demonstrated that DNN-based policies can predict generator dispatch decisions in real time under load and renewable uncertainty while producing near-optimal solutions and enforcing operational constraints \cite{li2024opf_surrogate}.

At the same time, power systems are becoming increasingly vulnerable to cyber and data-driven attacks. Data integrity attacks targeting OPF-related information can manipulate power dispatch solutions and compromise the operation of distributed OPF algorithms \cite{duan2018,yang2022}. In AC-OPF settings, falsification of active power, reactive power, and voltage information can alter dispatch outcomes and lead to economic or disruptive operational consequences \cite{yang2022}. As a result, identifying which input features most strongly influence OPF outcomes has emerged as an important research topic from both operational and security perspectives since important features could be the target of attacks. In this work, vulnerability analysis refers to identifying input features that can induce significant changes in OPF outcomes while remaining difficult to detect by security mechanisms. Unlike feature importance, which captures output sensitivity, vulnerability reflects a joint notion of operational impact and stealthiness under attack conditions. Prior studies have demonstrated that stealthy data-integrity attacks can
influence OPF outcomes while remaining undetected by conventional detection
mechanisms, highlighting the need for vulnerability analysis that jointly
considers operational impact and attack detectability
\cite{rahman2020,constanteflores2022}.

\subsection{Explainable AI for Power System Analysis}

To address the opacity of machine-learning-based power system models, explainable artificial intelligence (XAI) techniques have been proposed. Among these, Shapley value–based methods such as SHAP provide a principled framework for evaluating the contribution of individual features to model predictions based on cooperative game theory \cite{lundberg2017unified}. SHAP has been applied to a variety of power system applications, including
transient stability assessment, sensitivity analysis of active distribution
networks, reactive power control in photovoltaic systems, and transmission-line
risk diagnosis, demonstrating its usefulness in identifying influential system
variables and improving the interpretability of data-driven models
\cite{hamilton2024,lamprianidou2026,utama2022,deng2026}.

More recently, explainability methods have been applied to power system stability and OPF-related problems. Kilembe \textit{et al.} employed SHAP to analyze locational frequency stability and identify critical system variables influencing frequency-stability boundaries \cite{kilembe2025explainable}. Fritz and Bukhsh proposed counterfactual explanations for DC-OPF to reveal how minimal changes in input variables affect dispatch decisions \cite{fritz2025explainable}. In these studies, features with high explainability scores are often interpreted as critical or influential components of the system, as perturbing such features can significantly alter model outputs. Beyond conventional interpretability, XAI has also been incorporated into power-system cybersecurity applications to improve the transparency of false data injection attack detection and support attack localization \cite{dehbozorgi2025}.

\subsection{Limitations of Existing Vulnerability Analyses}

Despite their significance in modeling input-to-output relationship, existing explainability-based analyses suffer from a fundamental limitation when interpreted from the cybersecurity perspective. Most prior works focus exclusively on \emph{input-to-output} relationships, identifying features that strongly influence OPF outputs. However, practical security and anomaly detection mechanisms often operate by monitoring input-to-input relationships, identifying deviations from learned correlations and normal operating patterns using data-driven models such as autoencoders and one-class classifiers \cite{zhou2017,feng2024,roy2022}. Autoencoder-based models have been increasingly applied in power-system cybersecurity and anomaly detection to learn normal operating patterns and identify deviations associated with cyber or physical anomalies \cite{feng2024,tabassum2024}.

As a consequence, features that are operationally important may still be easily detectable when perturbed, while features with lower operational sensitivity may allow stealthier but still effective manipulation. This mismatch implies that feature importance derived solely from OPF surrogate explainability does not necessarily correspond to vulnerability importance under realistic security settings. To date, existing explainability-based vulnerability analyses in power system have not systematically incorporated detectability considerations arising from input-to-input security modeling.

\subsection{Problem Statement}

From an attacker's perspective, vulnerability analysis for OPF inputs must balance two competing objectives: inducing significant changes in OPF outputs while remaining undetected by attack detection mechanisms. In practice, attackers rarely possess knowledge of the specific detection models deployed by system operators, which may include autoencoder-based detectors, one-class support vector machines (SVMs), ensemble anomaly detection methods, or others.

This leads to the following key research questions:
\begin{itemize}
    \item Are features identified as important by OPF surrogate explainability methods truly vulnerable when an attack detection mechanism is in place?
    \item How can vulnerability be quantified when attack detection mechanisms rely on input-to-input relationships rather than OPF outputs?
    \item Can vulnerable features be identified without assuming knowledge of the underlying attack-detection model?
\end{itemize}

Addressing these questions requires a vulnerability definition that jointly accounts for operational impact and detectability under unknown attack detection mechanisms.

\subsection{Contributions}

The main contributions of this work are summarized as follows:
\begin{itemize}
    \item To the best of our knowledge, we propose the first defense-aware explainable AI framework for OPF vulnerability analysis that jointly models operational impact and feature-level detectability.
    
    \item We introduce a novel feature-level vulnerability metric by combining OPF-based SHAP values with detectability-oriented SHAP values derived from a vanilla autoencoder. The key goal here is to extend traditional input--output explainability toward a defense-aware notion of vulnerability through the incorporation of input--input relationship modeling.
    
    \item To evaluate the effectiveness of the proposed framework, we construct independent reference vulnerability rankings using a diverse set of security and anomaly detection models, including autoencoder-based detectors, linear one-class SVM, and Isolation Forest, under realistic attacker assumptions.
    
    \item We validate the proposed framework on IEEE 30-bus, 57-bus, and 118-bus systems using a realistic synthetic dataset generated by solving AC-OPF formulations, demonstrating consistent and scalable improvement over OPF-only explainability methods across varying system sizes and security models.
\end{itemize}

\subsection{Paper Organization}

The remainder of the paper is organized as follows. Section~II presents a brief overview of our work. Section~III describes details of the defense-aware vulnerability analysis framework. Section~IV introduces our approach for generating reference vulnerability rankings. Section~V details the experimental setup and evaluation protocol. Section~VI presents the experiment results and analysis. Section VII discusses the limitations and future research directions, while Section VIII concludes the paper.
\section{Overview}

In this research, our goal is to develop a defense-aware explainable AI framework for vulnerability analysis that integrates a machine-learning-based OPF surrogate model with an autoencoder-based security representation. The key motivation behind jointly using an OPF surrogate and a vanilla autoencoder is to explicitly capture both \emph{input-to-output} relationships, which describe how system inputs affect OPF solutions, and \emph{input-to-input} relationships, which are used by security and anomaly detection mechanisms. The framework is illustrated in Fig. 1.

First, power system optimal power flow data are used to train a neural network surrogate model that approximates the AC-OPF mapping from system inputs (e.g., bus load demands) to optimal operating variables (e.g., generator setpoints). Explainable AI techniques based on Shapley values (SHAP) are then applied to this surrogate model to quantify the importance of each input feature from the output perspective, i.e., how strongly each input influences the OPF solution.
Next, using the same input data, we train a vanilla autoencoder to learn the underlying input-to-input dependencies present in normal system operation. The autoencoder captures correlations and structural relationships among input features that are typically leveraged by security and anomaly detection systems. SHAP analysis is subsequently applied to the autoencoder model to compute Shapley values for all input features, which reflect the contribution of each feature to the autoencoder’s reconstruction behavior and, by extension, its detectability.

Here, a vanilla autoencoder is adopted because it explicitly learns input--input relationships that characterize normal operating behavior, which closely aligns with how practical security and anomaly detection mechanisms function. In contrast, classical models such as one-class SVMs primarily learn global decision boundaries over the input distribution and do not model feature-wise dependencies. Moreover, the reconstruction-based objective of autoencoders enables stable and interpretable feature-level detectability attribution via SHAP, which is essential for the proposed vulnerability analysis.

To jointly account for operational influence and detectability, we define a vulnerability metric based on the ratio of the Shapley values obtained from the OPF surrogate model and the vanilla autoencoder model. Intuitively, this ratio highlights features that have a strong impact on OPF outputs while exhibiting relatively weak detectability under the learned input-to-input relationships. Features with higher ratio values are therefore considered more vulnerable from the security perspective. 

\begin{figure}[t]
    \centering
    \includegraphics[width=1.2\columnwidth]{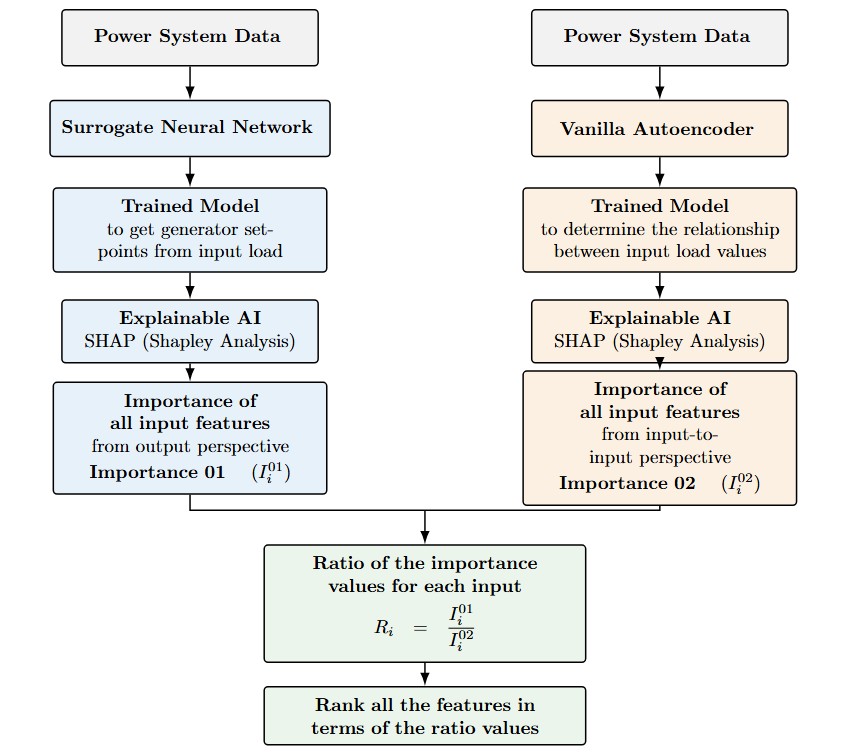}
    \caption{Proposed Defense-Aware Feature Vulnerability Analysis Framework.}
    \label{fig:framework}
\end{figure}

To evaluate the effectiveness of the proposed framework, reference vulnerability rankings (which will serve as the benchmark) are generated using multiple attack detection models. For each model, individual input features are perturbed under controlled magnitudes, and the resulting system response is assessed based on two criteria: i) whether the perturbation remains undetected by the attack-detection model, and ii) the extent to which it alters the OPF outputs. Features that induce large output deviations while consistently evading detection are ranked as more vulnerable under the corresponding security structure. This procedure is repeated across all considered attack-detection models to obtain multiple reference rankings, which are subsequently compared with vulnerability rankings derived from the proposed ratio-based metric to quantitatively assess the effectiveness of our framework.
\section{Defense-Aware Explainable AI Framework}

\subsection{Problem Setup}

Power system is a complex network with many components such as buses, generators, and transmission lines, where system operation is governed by the alternating-current optimal power flow (AC-OPF) problem. The AC-OPF formulation considered in this work follows the standard nonlinear
AC power-flow equations and operational constraints \cite{wood2012power}. Let $\mathbf{x} \in \mathbb{R}^{d}$ denote the vector of system input features, where $\mathbf{x}$ typically consists of active and reactive power demands at load buses. Let $\mathbf{y} \in \mathbb{R}^{m}$ denote the vector of OPF outputs, which may include generator active and reactive power set-points, bus voltage magnitudes, and other operational variables of interest. In this analysis, we restrict the OPF output vector to generator active and reactive power set-points and voltage magnitudes at generator buses, as these variables are primary OPF decision variables and directly represent the control actions implemented by system operators.

Given an input $\mathbf{x}$, the AC-OPF problem determines the optimal operating point $\mathbf{y}$ by solving a nonlinear, nonconvex optimization problem subject to power flow equations, generator limits, voltage constraints, and network constraints. We denote the mapping induced by the OPF solver as
\begin{equation}
\mathbf{y} = f_{\text{OPF}}(\mathbf{x}).
\end{equation}

Using a synthetic power system dataset, we construct a collection of input–output pairs
\begin{equation}
\mathcal{D} = \{ (\mathbf{x}_i, \mathbf{y}_i) \}_{i=1}^{N},
\end{equation}
where $\mathbf{x}_i \in \mathbb{R}^{d}$ denotes the observed system input features and $\mathbf{y}_i \in \mathbb{R}^{m}$ denotes the corresponding generator set-points obtained from operational data.  All models are trained and evaluated using these measurements, ensuring that the proposed vulnerability analysis reflects realistic operating conditions.

In this work, vulnerability analysis is performed at the level of input features. We use the term \emph{feature importance} to refer to the sensitivity of OPF outputs with respect to individual input features, as quantified through explainable AI methods. In contrast, we use the term \emph{vulnerability} to describe the susceptibility of an input feature to induce significant changes in OPF outputs while remaining difficult to detect by a security or anomaly detection mechanism. This distinction is central to the proposed framework and is formalized in subsequent sections.

Throughout the paper, bold lowercase letters denote vectors, and subscripts are used to index individual features or data samples. Unless otherwise stated, all models are trained using data corresponding to normal system operation.

\subsection{Machine-Learning-Based OPF Surrogate Model}

The surrogate model that maps system load inputs to OPF generator set-points is implemented as a compact feedforward neural network with two hidden layers and ReLU activations. The model takes active and reactive load demands as inputs and predicts generator active power, reactive power, and voltage magnitudes at all generator buses.

The network is composed of successive linear layers followed by nonlinear activation functions, which allows it to learn a flexible mapping from inputs to outputs without introducing excessive architectural complexity. This formulation is sufficient to capture the input–output relationships present in the training data while remaining easy to train and evaluate.

In addition, shallow architectures reduce inference latency and sensitivity to small input perturbations, which is desirable when vulnerability assessment must be performed repeatedly across large datasets and operating conditions.

Early stopping based on validation performance is used to regularize training and promote generalization, further reinforcing the stability of learned sensitivities. Overall, this modeling approach provides an effective balance between accuracy, robustness, and interpretability, making it well suited for explainability-driven OPF vulnerability and security analyses.

\subsection{Feature Importance in OPF via SHAP}

To quantify the influence of individual input features on OPF outcomes, we employ a Shapley value–based explainable AI approach (SHAP) on the OPF surrogate model.

Given a trained surrogate model $\hat{f}_{\text{OPF}}(\cdot)$, SHAP attributes the model prediction to individual input features by computing Shapley values derived from cooperative game theory. For the $i$-th input feature, the corresponding Shapley value $\phi_i^{\text{OPF}}$ is defined as
\begin{equation}
\phi_i^{\text{OPF}} =
\sum_{S \subseteq \mathcal{F} \setminus \{i\}}
\frac{|S|!\,(|\mathcal{F}| - |S| - 1)!}{|\mathcal{F}|!}
\begin{aligned}[t]
\Big[&\hat{f}_{\text{OPF}}(\mathbf{x}_{S \cup \{i\}}) \\
     &- \hat{f}_{\text{OPF}}(\mathbf{x}_{S})\Big]
\end{aligned}
\end{equation}
where $\mathcal{F} = \{1,2,\dots,d\}$ denotes the set of all input features, and $\mathbf{x}_{S}$ represents the input vector in which only features indexed by the subset $S$ are present.

Since the OPF surrogate produces multiple generator set-point outputs, SHAP values are computed for each output dimension and subsequently aggregated to obtain a single importance score per input feature. Specifically, the aggregated OPF-based SHAP value for feature $i$ is given by
\begin{equation}
\phi_i^{\text{OPF}} = \frac{1}{m} \sum_{j=1}^{m} \left| \phi_{i,j}^{\text{OPF}} \right|,
\end{equation}
where $\phi_{i,j}^{\text{OPF}}$ denotes the Shapley value of the $i$-th input feature with respect to the $j$-th generator set-point output.

The resulting OPF-based SHAP values, $\phi_i^{\text{OPF}}$, capture the operational influence of each input feature on generator set-point decisions. Features with larger $\phi_i^{\text{OPF}}$ values exert a stronger effect on OPF outputs and are therefore considered more important from an output sensitivity perspective. However, these importance scores do not account for detectability by security mechanisms and should not be interpreted as vulnerability measures. This limitation motivates the incorporation of defense-aware modeling in subsequent sections.

\subsection{Modeling Feature-Level Detectability via Input-to-Input Relationships}

In practical power system operations, analyses such as optimal power flow (OPF) are typically protected by security or anomaly detection mechanisms deployed by system operators. Unlike OPF models that focus on \emph{input-to-output} relationships, these security mechanisms primarily monitor \emph{input-to-input} relationships within the measurement data to identify abnormal or malicious behavior. Such models assess whether observed input patterns conform to learned correlations and structural dependencies among features under normal operating conditions.

A key observation motivating this work is that perturbations applied to different input features do not affect these learned relationships uniformly. Perturbing certain features can significantly disrupt the reconstructed input patterns, making such changes easily detectable by attack detection models, whereas perturbations applied to other features may have a relatively minor impact on reconstruction behavior and may remain stealthy. This variability arises from the underlying input-to-input dependencies present in the data and directly influences the vulnerability of individual features from the attack perspective.

To capture this behavior, we employ a vanilla autoencoder as a representative model for learning input-to-input relationships. The autoencoder consists of an encoder function $g(\cdot)$ and a decoder function $h(\cdot)$, which together produce a reconstruction,
\begin{equation}
\hat{\mathbf{x}} = h\big(g(\mathbf{x})\big).
\end{equation}
The model is trained using only normal operating data by minimizing the reconstruction loss
\begin{equation}
\mathcal{L}_{\text{AE}} =
\mathbb{E}_{\mathbf{x}}
\left[
\left\|
\mathbf{x} - \hat{\mathbf{x}}
\right\|_2^2
\right].
\end{equation}

While reconstruction error provides a coarse indication of detectability, it does not directly reveal the contribution of individual input features to the detection process. To obtain feature-level insights from a detectability perspective, we apply Shapley value–based explainable AI techniques (SHAP) to the trained autoencoder. Specifically, SHAP values are computed with respect to the autoencoder’s reconstruction behavior to quantify how each input feature contributes to maintaining or violating the learned input-to-input relationships.

The resulting autoencoder-based SHAP values provide a measure of feature importance from a \emph{detectability} standpoint. Features with larger SHAP values strongly influence the reconstruction process and are therefore more likely to trigger detection when perturbed, whereas features with smaller SHAP values have weaker influence on the learned input-to-input relationships and may allow perturbations to remain stealthy.

\subsection{Vulnerability Metric Construction}

Based on the preceding analyses, vulnerability in the proposed framework is defined as a joint function of operational impact and detectability. OPF-based SHAP values quantify how strongly perturbations to individual input features influence generator set-point decisions, while autoencoder-based SHAP values quantify how sensitive these perturbations are to attack detection mechanisms. From the attack perspective, a feature is considered vulnerable if it can induce significant changes in OPF outputs while exhibiting low detectability under attack detection settings.

To capture this interaction, we define a defense-aware vulnerability metric by combining the two SHAP-based measures. Specifically, for the $i$-th input feature, the vulnerability score is defined as
\begin{equation}
V_i = \frac{\phi_i^{\text{OPF}}}{\phi_i^{\text{AE}}},
\end{equation}
where $\phi_i^{\text{OPF}}$ denotes the OPF-based SHAP value representing operational influence, and $\phi_i^{\text{AE}}$ denotes the autoencoder-based SHAP value representing detectability. 

The resulting vulnerability scores $V_i$ provide a ranking of input features from the attack perspective. Features with higher $V_i$ values represent more promising attack surfaces, as perturbations applied to these features are more likely to remain stealthy while causing substantial operational impact. By explicitly incorporating both operational sensitivity and detectability, the proposed metric extends traditional explainability-based importance measures toward a defense-aware notion of vulnerability for the OPF problem.

\section{Reference Vulnerability Ranking Generation}

To validate the proposed framework, it is necessary to assess whether the features identified as vulnerable by the framework indeed exhibit vulnerability from an empirical security perspective. This requires comparison against independently generated reference vulnerability rankings. In this section, we describe the procedure used to generate these reference rankings, which serve as evaluation benchmarks for assessing the effectiveness of the proposed framework.

\subsection{Threat Model and Assumptions}

In this work, reference vulnerability rankings are generated from an attacker's perspective under realistic security assumptions. We consider an attacker who is capable of perturbing input measurements, such as load demands, but does not possess knowledge of the anomaly detection mechanisms deployed by the system operator. The design, structure, and parameters of the attack detection models are assumed to be controlled by the operator and remain unknown to the attacker.

The reference vulnerability ranking generation procedure is introduced solely for evaluation purposes and is not part of the proposed vulnerability identification framework. Its role is to provide an empirical benchmark against which the effectiveness of the proposed vulnerability metric can be assessed. This separation ensures that the evaluation does not introduce circular dependencies or implicitly favor the proposed framework.

Throughout this analysis, perturbations are restricted to bounded magnitudes to ensure physical plausibility and to avoid trivial detection. Specifically, perturbation bounds are selected based on empirical operating ranges observed in normal system data. In our experiments, we consider multiple physically consistent bounds for perturbation, including \([0,0.1]\), \([0,0.2]\), and \([0,0.5]\) per unit, which are consistent with the load measurements in the dataset.

While the absolute magnitude of OPF generator set-points deviation varies with the chosen perturbation bound, the relative ranking of vulnerable features remains largely unchanged across all tested bounds. This indicates that the proposed framework captures intrinsic feature-level vulnerability characteristics rather than artifacts of a particular perturbation scale.

The workflow to generate the reference vulnerability rankings is shown in Fig. 2.

\begin{figure}[t]
\centering
\includegraphics[width=1.15\columnwidth]{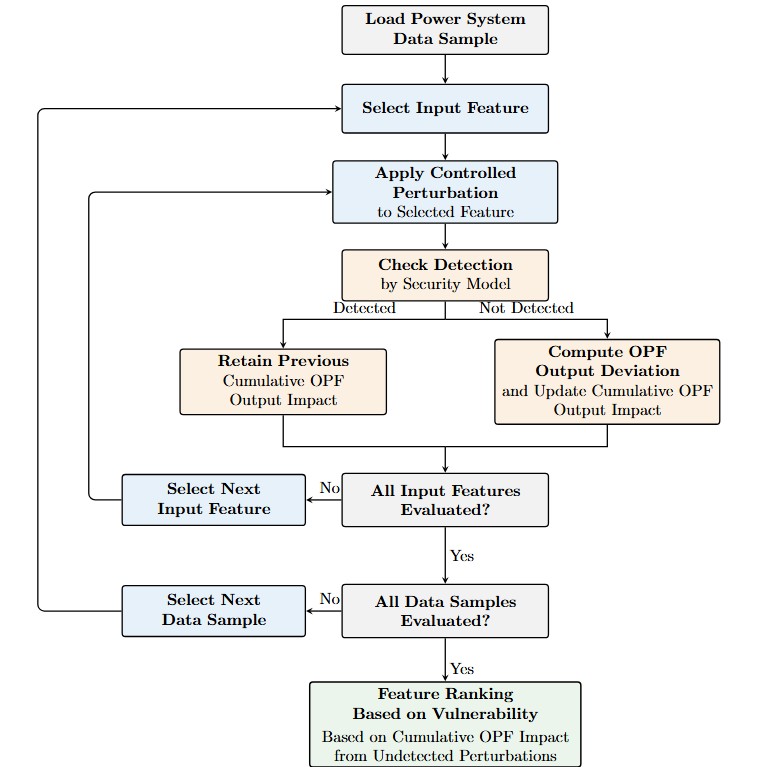}
\caption{Reference Vulnerability Ranking Generation Workflow}
\end{figure}

\subsection{Feature-Wise Perturbation Strategy}

To assess the vulnerability of individual input features, a feature-wise perturbation strategy is employed. For each input feature, perturbations are applied independently while all remaining features are held fixed. This one-feature-at-a-time approach allows us to determine the vulnerability contribution of each individual feature without conflating interactions among multiple inputs.

Perturbations are applied across multiple operating samples to account for variability in system conditions. For a given feature, the same perturbation procedure is implemented over all data samples in the dataset, and the resulting system responses are collected. This ensures that vulnerability assessments are not driven by isolated operating points but reflect consistent behavior across typical operating conditions.

By systematically perturbing each input feature under controlled and bounded conditions, this strategy enables a direct evaluation of how individual features influence detectability and operational impact when manipulated.

\subsection{Detection Models Considered}

To generate reference vulnerability rankings for validating the proposed framework, we evaluate feature-level perturbations under multiple anomaly detection models. These detection models are assumed to be deployed by the system operator to protect OPF analysis and are not known to the attacker. 

In this work, six representative attack or anomaly detection models are considered:

\begin{itemize}
    \item \textbf{Vanilla Autoencoder:} This model serves as a baseline security mechanism that captures general input-to-input correlations under normal operating conditions. It learns compact representations of input features and detects anomalies based on reconstruction behavior, making it a widely adopted reference for data-driven anomaly detection.
    
    \item \textbf{Contrastive Denoising Autoencoder:} This detection model is designed to be robust to measurement noise and small random perturbations. By learning to reconstruct clean inputs from corrupted versions, it emphasizes stable structural relationships among features and is effective at identifying coordinated deviations while tolerating benign noise.
    
    \item \textbf{Sparse Autoencoder:} This model enforces sparsity constraints on latent representations, encouraging selective activation of a limited number of latent units. As a result, it is sensitive to structured or coordinated perturbations that violate learned normal patterns, making it suitable for detecting subtle but systematic anomalies.
    
    \item \textbf{Adversarial Autoencoder:} This advanced security model imposes distributional constraints on latent representations by aligning them with a predefined prior distribution. During training, the model jointly learns input reconstruction and latent distribution matching. When manipulated or abnormal inputs are introduced, their latent representations deviate from the learned prior, enabling effective detection of anomalous behavior.
    \item \textbf{Linear One-Class Support Vector Machine (OC-SVM):} 
    The linear OC-SVM learns a hyperplane that separates normal operating data from the origin in the input feature space with maximum margin. By using a linear kernel, the model captures dominant linear relationships among input features without introducing nonlinear transformations. During training, the OC-SVM identifies a boundary that characterizes the distribution of normal data based on these linear dependencies. When manipulated or abnormal inputs are introduced, they violate the learned relationships and are mapped outside the decision boundary, leading to anomaly detection. The linear formulation makes this model particularly sensitive to structured deviations that disrupt global feature consistency, while remaining computationally efficient and interpretable.

    \item \textbf{Isolation Forest (IF):} This anomaly detection model identifies abnormal inputs by isolating data points through random feature partitioning. The model exploits the observation that anomalous samples are easier to separate from normal data and therefore require fewer partitioning steps. During operation, inputs that are isolated with shorter path lengths are classified as anomalous. Isolation Forest does not rely on distance or density assumptions and is effective at detecting irregular or rare feature combinations arising from coordinated or unstructured perturbations.
\end{itemize}

For a given detection model, input features are perturbed in a controlled manner, and only those perturbations that bypass the detection mechanism are considered successful from a stealth perspective. For such undetected perturbations, the resulting changes in OPF outputs, specifically generator set-points, are recorded and analyzed. By aggregating these changes in outputs across all operating samples in reference subset, we obtain the total impacts of all individual feature perturbations which are used for reference vulnerability rankings generation. To efficiently identify the maximum perturbation magnitude that remains undetected, we employ a binary search strategy over a bounded perturbation interval. A binary search procedure is used to efficiently determine the maximum perturbation magnitude that remains undetected. Starting from an initial interval $[0, \delta_{\text{init}}]$, the method iteratively refines the perturbation by halving the search range based on whether the perturbed input satisfies the stealth constraint. This enables precise estimation of the largest admissible perturbation with logarithmic complexity. Algorithm 1 shows the reference vulnerability ranking generation process for each detection model.

\begin{algorithm}[t]
\caption{Reference Vulnerability Ranking Generation}
\label{alg:stealth_vulnerability}
\small
\begin{algorithmic}[1]

\STATE \textbf{Input:}
\STATE \hspace{1em} Reference dataset $\mathcal{D}_{ref} = \{x^{(n)}\}_{n=1}^{N}$
\STATE \hspace{1em} OPF surrogate model $f_{\text{OPF}}(\cdot)$
\STATE \hspace{1em} Attack detector $g(\cdot)$
\STATE \hspace{1em} Stealth threshold $\tau$
\STATE \hspace{1em} Perturbation bound $\delta_{\text{init}}$
\STATE \hspace{1em} Binary search iterations $T$

\STATE \textbf{Output:}
\STATE \hspace{1em} Feature vulnerability scores $\{V_j\}_{j=1}^{d}$

\STATE Initialize $V_j \leftarrow 0$ for all $j=1,\dots,d$

\FOR{each $x \in \mathcal{D}_{ref}$}
    \STATE $y_{\text{base}} \leftarrow f_{\text{OPF}}(x)$
    
    \FOR{$j = 1$ to $d$}
        \STATE $\delta_{\max} \leftarrow 0$, $s^* \leftarrow 0$
        
        \FOR{$s \in \{-1, +1\}$}
            \STATE $low \leftarrow 0$, $high \leftarrow \delta_{\text{init}}$
            
            \FOR{$t = 1$ to $T$}
                \STATE $\delta \leftarrow (low + high)/2$
                \STATE $\tilde{x} \leftarrow x$, $\tilde{x}_j \leftarrow x_j + s \cdot \delta$
                
                \IF{$g(\tilde{x})$ satisfies $\tau$}
                    \STATE $low \leftarrow \delta$
                    
                    \IF{$\delta > \delta_{\max}$}
                        \STATE $\delta_{\max} \leftarrow \delta$
                        \STATE $s^* \leftarrow s$
                    \ENDIF
                \ELSE
                    \STATE $high \leftarrow \delta$
                \ENDIF
            \ENDFOR
        \ENDFOR
        
        \IF{$\delta_{\max} = 0$}
            \STATE \textbf{continue}
        \ENDIF
        
        \STATE $x^* \leftarrow x$, $x^*_j \leftarrow x_j + s^* \cdot \delta_{\max}$
        \STATE $y_{\text{attack}} \leftarrow f_{\text{OPF}}(x^*)$
        \STATE $\Delta \leftarrow \|y_{\text{attack}} - y_{\text{base}}\|_1$
        
        \STATE $V_j \leftarrow V_j + \Delta$
        
    \ENDFOR
\ENDFOR

\STATE \textbf{return} features ranked by $V_j$ (descending)

\end{algorithmic}
\end{algorithm}
\section{Experimental Setup}

This section describes the dataset, system configurations, model training procedures, and evaluation protocol used to assess the proposed vulnerability analysis framework. All experiments are conducted using a synthetic power system AC-OPF dataset.

\subsection{Dataset Description}

All experiments in this work are based on a publicly available synthetic
power system dataset released by the National Laboratory of the Rockies
\cite{joswig2021opflearn}. The dataset contains operational measurements from realistic power system scenarios, including load demands and generator operating points obtained from AC-OPF solutions.

Input features are constructed using active and reactive load demands at selected buses, while output variables correspond to generator active and reactive power, and voltage set-points. All models are trained and evaluated using measurements drawn directly from the dataset. This ensures that both the surrogate OPF model and the detection models learn operating patterns from the same origin.

\subsection{System Configurations}

The proposed framework is evaluated on three benchmark IEEE test systems covered by the dataset: the IEEE 30-bus, IEEE 57-bus, and IEEE 118-bus systems. These systems represent increasing levels of network complexity and dimensionality. For each system, the corresponding subset of load and generator variables is extracted from the dataset to form system-specific input–output pairs.

The IEEE 30-bus system includes a moderate number of load features and generator set-points, while the IEEE 57-bus and IEEE 118-bus systems progressively increase the dimensionality of the input space. This setup allows assessment of the scalability of the proposed vulnerability metric under realistic operating conditions.

\subsection{Model Training and Evaluation Protocol}

For each IEEE test system, the dataset is partitioned into three disjoint subsets consisting of 4,000 samples for training, 1,000 samples for validation, and 5,000 samples reserved exclusively for reference vulnerability ranking generation. The proposed feature ranking is obtained during the training and validation stage using the trained models and their corresponding feature-attribution results. The held-out 5,000-sample subset is then used to independently generate the reference vulnerability ranking, allowing us to evaluate how well the proposed ranking agrees with vulnerability behavior observed on unseen operating samples. This partitioning ensures a clear separation between model training and vulnerability evaluation, thereby avoiding data leakage and evaluation bias.

The OPF surrogate model is trained using the training subset to approximate the mapping between load demands and generator set-points. Model parameters are optimized by minimizing the mean squared error loss. The validation subset is used for early stopping to prevent overfitting, and training is terminated once validation performance stabilizes. Prediction accuracy on the validation set is used to confirm that the surrogate provides a faithful approximation of OPF behavior prior to explainability and vulnerability analysis.

All security and anomaly detection models, including autoencoder-based detectors, linear one-class SVM, and Isolation Forest, are trained exclusively using normal operating data from the training subset. Autoencoder models are optimized to reconstruct input measurements and learn characteristic input-to-input relationships under nominal conditions, with early stopping applied based on validation performance. No attack data or perturbed samples are used during training or validation of the security models.

Vulnerability evaluation is performed using the dedicated reference subset consisting of 5{,}000 samples for each system. This choice reflects the objective of the study, which is to assess feature-level vulnerability rankings rather than predictive generalization. Using a separate reference set ensures an independent and unbiased evaluation of vulnerability behavior.

The reference vulnerability ranking generation process is performed across different configurations and parameter settings of each detection model to account for variability in detection behavior and decision boundaries. For each detection model, reference vulnerability rankings are generated independently for each configuration. The ranking generated by our solution is then compared with each configuration-specific reference ranking by measuring feature overlap in the top 10\% and top 20\% of the ranked features. The resulting similarity scores are averaged across configurations to obtain a final match percentage for that detection model. 

Note that this averaging is performed only on similarity scores, while the reference rankings themselves are not aggregated across configurations or across different security models. Each reference ranking therefore reflects the vulnerability structure induced by a specific detection mechanism and setting. Importantly, these reference rankings are generated independently of the proposed vulnerability metric and are used solely as empirical benchmarks for evaluation, ensuring a fair and unbiased comparison.

The OPF surrogate model’s SHAP-based feature importance serves as the baseline, representing conventional input--output explainability in OPF analysis. Comparing this baseline with the proposed defense-aware framework enables direct evaluation of the impact of incorporating detectability into vulnerability identification.
\section{Results and Analysis}

This section evaluates the proposed vulnerability analysis framework by comparing its feature rankings with a conventional baseline derived from OPF surrogate model SHAP importance (denoted as \emph{OPF-SHAP}). For each detection model, we report the similarity of the top-ranked features identified by (i) the OPF-SHAP baseline and (ii) the proposed framework to the corresponding reference vulnerability ranking.

\begin{table*}[!t]
\centering

\caption{Similarity Comparison for IEEE 30-Bus System (Top 10\% Features)}
\label{tab:30_top10}
\begin{tabular}{lcccccc}
\toprule
\textbf{Method} & \textbf{Vanilla AE} & \textbf{CDAE} & \textbf{Sparse AE} & \textbf{Adv. AE} & \textbf{SVM} & \textbf{IF} \\
\midrule
OPF-SHAP & 20\% & 40\% & 40\% & 60\% & 60\% & 60\% \\
Proposed Framework & 80\% & 80\% & 80\% & 80\% & 80\% & 80\% \\
\bottomrule
\end{tabular}

\vspace{0.5em}

\caption{Similarity Comparison for IEEE 30-Bus System (Top 20\% Features)}
\label{tab:30_top20}
\begin{tabular}{lcccccc}
\toprule
\textbf{Method} & \textbf{Vanilla AE} & \textbf{CDAE} & \textbf{Sparse AE} & \textbf{Adv. AE} & \textbf{SVM} & \textbf{IF} \\
\midrule
OPF-SHAP & 66.67\% & 66.67\% & 55.56\% & 66.67\% & 66.67\% & 55.56\% \\
Proposed Framework & 88.89\% & 88.89\% & 88.89\% & 88.89\% & 77.78\% & 88.89\% \\
\bottomrule
\end{tabular}

\vspace{0.5em}

\caption{Similarity Comparison for IEEE 57-Bus System (Top 10\% Features)}
\label{tab:57_top10}
\begin{tabular}{lcccccc}
\toprule
\textbf{Method} & \textbf{Vanilla AE} & \textbf{CDAE} & \textbf{Sparse AE} & \textbf{Adv. AE} & \textbf{SVM} & \textbf{IF} \\
\midrule
OPF-SHAP & 22.22\% & 22.22\% & 22.22\% & 33.33\% & 22.22\% & 22.22\% \\
Proposed Framework & 88.89\% & 88.89\% & 88.89\% & 88.89\% & 77.78\% & 88.89\% \\
\bottomrule
\end{tabular}

\vspace{0.5em}

\caption{Similarity Comparison for IEEE 57-Bus System (Top 20\% Features)}
\label{tab:57_top20}
\begin{tabular}{lcccccc}
\toprule
\textbf{Method} & \textbf{Vanilla AE} & \textbf{CDAE} & \textbf{Sparse AE} & \textbf{Adv. AE} & \textbf{SVM} & \textbf{IF} \\
\midrule
OPF-SHAP & 35.29\% & 41.18\% & 29.41\% & 29.41\% & 29.41\% & 29.41\% \\
Proposed Framework & 94.11\% & 88.24\% & 88.24\% & 88.24\% & 88.24\% & 82.35\% \\
\bottomrule
\end{tabular}

\vspace{0.5em}

\caption{Similarity Comparison for IEEE 118-Bus System (Top 10\% Features)}
\label{tab:118_top10}
\begin{tabular}{lcccccc}
\toprule
\textbf{Method} & \textbf{Vanilla AE} & \textbf{CDAE} & \textbf{Sparse AE} & \textbf{Adv. AE} & \textbf{SVM} & \textbf{IF} \\
\midrule
OPF-SHAP & 20\% & 20\% & 20\% & 25\% & 20\% & 20\% \\
Proposed Framework & 70\% & 75\% & 75\% & 75\% & 70\% & 75\% \\
\bottomrule
\end{tabular}

\vspace{0.5em}

\caption{Similarity Comparison for IEEE 118-Bus System (Top 20\% Features)}
\label{tab:118_top20}
\begin{tabular}{lcccccc}
\toprule
\textbf{Method} & \textbf{Vanilla AE} & \textbf{CDAE} & \textbf{Sparse AE} & \textbf{Adv. AE} & \textbf{SVM} & \textbf{IF} \\
\midrule
OPF-SHAP & 25\% & 20\% & 25\% & 25\% & 27.5\% & 22.5\% \\
Proposed Framework & 75\% & 75\% & 77.5\% & 77.5\% & 72.5\% & 75\% \\
\bottomrule
\end{tabular}

\end{table*}

\subsection{IEEE 30-Bus System}

For the IEEE 30-bus system, the dataset contains 44 input features and 18 OPF outputs (generator set-points). Table~\ref{tab:30_top10} and Table~\ref{tab:30_top20} report the similarity of the baseline and our solution with the reference rankings under six attack detection models.

For the IEEE 30-bus system, the baseline OPF-SHAP ranking shows moderate but inconsistent agreement with the reference vulnerability rankings. As shown in Table~\ref{tab:30_top10}, the overlap between OPF-SHAP and the reference rankings ranges from 20\% to 60\% across different detection models. This indicates that although OPF-SHAP captures features that strongly influence generator set-points, these features are not consistently vulnerable when detectability constraints are considered.

In contrast, the proposed framework consistently identifies vulnerable features across all six detection models, achieving an 80\% overlap in the top-10\% feature set. This consistency demonstrates that incorporating both operational impact and detectability leads to more reliable vulnerability identification.

At the top-20\% level (Table~\ref{tab:30_top20}), both approaches exhibit improved overlap due to the relaxed selection threshold. However, the proposed framework continues to outperform the baseline across all detection models. These results highlight the advantage of the proposed framework in identifying a small and reliable set of vulnerable features.

\subsection{IEEE 57-Bus System}

For the IEEE 57-bus system, the dataset contains 84 input features and 21 OPF outputs. Table~\ref{tab:57_top10} and Table~\ref{tab:57_top20} summarize the similarity results under the same six attack detection models.

For the IEEE 57-bus system, the limitations of OPF-only explainability become more pronounced. The OPF-SHAP ranking exhibits relatively low agreement with the reference vulnerability rankings, achieving less than 35\% overlap in the top-10\% feature set and below 42\% overlap in the top-20\% feature set across all detection models.

The proposed framework substantially improves vulnerability identification in this higher-dimensional setting. Across all detection models, it consistently achieves high overlap with the reference rankings, often exceeding 85\% in the top-20\% regime. This improvement demonstrates that detectability information becomes increasingly important as the number of input features grows.

Overall, these results confirm that identifying vulnerable features based solely on input-to-output sensitivity is insufficient for medium-scale systems, whereas the proposed framework remains effective and robust.
\subsection{IEEE 118-Bus System}

For the IEEE 118-bus system, the dataset contains 198 input features and 162 OPF outputs. Results are summarized in Table~\ref{tab:118_top10} and Table~\ref{tab:118_top20}.

The IEEE 118-bus system represents the most challenging scenario due to its high input dimensionality. In this case, OPF-SHAP exhibits consistently low agreement with the reference vulnerability rankings, with overlap remaining below 30\% across all security models and both selection thresholds. This indicates that operational importance alone fails to capture vulnerability in large-scale systems.

In contrast, the proposed framework maintains strong and stable performance across all evaluated attack detection models. It achieves 70--77.5\% overlap with the reference rankings in both the top-10\% and top-20\% feature sets. Notably, this performance is consistent across reconstruction-based detectors as well as classical anomaly detection models such as one-class SVM and Isolation Forest.

These results demonstrate that as system size increases, the gap between feature importance and true vulnerability widens, reinforcing the necessity of defense-aware vulnerability analysis. 

\subsection{Remarks}

The results across all IEEE test systems demonstrate a clear distinction between feature importance derived from input-to-output sensitivity and feature vulnerability under security constraints. While OPF surrogate-based SHAP analysis identifies features that strongly influence generator set-points, it does not account for detectability by attack detection models that monitor consistency among input measurements. As a result, OPF-only explainability frequently highlights features that are operationally important but do not constitute realistic attack surfaces from a security perspective.

The proposed framework addresses this limitation by explicitly incorporating input--input relationship modeling through a vanilla autoencoder. By combining OPF-based SHAP values with detectability-oriented SHAP values derived from the autoencoder, the resulting vulnerability metric prioritizes features that simultaneously induce significant operational impact and remain stealthy under security measures. This distinction is particularly evident in the strict top-10\% regime, where identifying a small set of truly vulnerable features is most critical for attack generation and defense evaluation.

An important observation from the results is the robustness of the proposed framework across diverse security models. Despite variations in detection mechanisms, including denoising autoencoders, sparsity-enforcing models, adversarial constraints, and classical anomaly detection approaches such as one-class SVM and Isolation Forest, the proposed vulnerability rankings maintain strong agreement with independently generated reference rankings. This consistency suggests that modeling input--input relationships using a vanilla autoencoder captures fundamental detectability characteristics that generalize across a broad class of attack detection systems, even when the attacker lacks knowledge of the exact defense structure.

Finally, the results highlight that vulnerability assessment becomes increasingly challenging as system dimensionality grows. For medium-scale and large-scale systems such as the IEEE 57-bus and IEEE 118-bus networks, the effectiveness of OPF-only explainability diminishes as operational sensitivity and detectability become increasingly decoupled. In contrast, the proposed framework maintains strong performance across all evaluated system sizes, demonstrating that defense-aware modeling is essential for scalable and realistic vulnerability analysis in modern power systems.

\section{Limitations and Future Work}

While the proposed framework provides a principled and effective approach for feature-level vulnerability analysis in optimal power flow problems, several limitations remain and motivate future research directions.

First, the reference vulnerability ranking generation procedure considers feature-wise perturbations applied independently. Although this allows clear attribution of vulnerability to individual input features, coordinated multi-feature perturbations are not explicitly modeled. Extending the framework to analyze joint or correlated perturbations would enable the study of more sophisticated attack strategies.

Moreover, future work can extend the evaluation to additional anomaly detection architectures and security mechanisms to further assess the generalizability of the proposed vulnerability ranking.

Despite these limitations, the proposed framework establishes a foundation for defense-aware explainable vulnerability analysis in power systems and opens multiple avenues for advancing attack modeling, defense evaluation, and system hardening.
\section{Conclusion}

This work presents a defense-aware explainable AI framework for feature-level vulnerability analysis in optimal power flow problems. By distinguishing feature importance from vulnerability and explicitly accounting for detectability through input--input relationship modeling, the proposed approach overcomes key limitations of conventional OPF-only explainability. Extensive evaluation across multiple IEEE test systems and attack detection models demonstrates that the proposed framework more accurately identifies attack-relevant vulnerable features than traditional SHAP-based importance analysis. The results highlight the necessity of incorporating defense-aware modeling into explainable AI frameworks for power system vulnerability assessment.
\FloatBarrier

\end{document}